\documentclass[letterpaper, 10 pt, conference]{ieeeconf}  % Comment this line out
\IEEEoverridecommandlockouts                              % This command is only
\usepackage{graphicx}
\usepackage{float}
\usepackage{booktabs}
\usepackage{array}
\usepackage{amsmath} 
\usepackage{url}
\usepackage{setspace}
\usepackage{tikz}
\usepackage{lipsum}
\usepackage{comment}
\usepackage{amssymb}
\usepackage{multirow}
\usepackage{cite}
\usepackage{siunitx}
\usepackage{cleveref}
\let\labelindent\relax
\usepackage{enumitem}

\usepackage{amsthm}

\title{\LARGE \bf
Robust State of Health Estimation of Lithium-ion Batteries Using Convolutional Neural Network and Random Forest
}

\author{Niankai Yang$^{1}$, Ziyou Song$^{2}$, Heath Hofmann$^{2}$, and Jing Sun$^{1}$  

\thanks{$^*$This paper is based upon the work supported by the U.S. Office of Naval Research (ONR) under Grants N00014-16-1-3108 and N00014-18-2330}
\thanks{$^{1}$Niankai Yang and Jing Sun are with Department of Naval Architecture and Marine Engineering, University of Michigan, Ann Arbor, MI 48109, USA. (e-mail: \{ynk,\;jingsun\}@umich.edu)}
\thanks{$^{2}$Ziyou Song and Heath Hofmann are with Department of Electrical Engineering and Computer Science, University of Michigan, Ann Arbor, MI 48109, USA. (e-mail: \{ziyou,\;hofmann\}@umich.edu)}% ,   
}
\begin{document}

\maketitle

%%%%%%%%%%%%%%%%%%%%%%%%%%%%%%%%%%%%%%%%%%%%%%%%%%%%%%%%%%%%%%%%%%%%%%%%%%%%%%%%%%%%
\begin{abstract}
The State of Health (SOH) of lithium-ion batteries is directly related to their safety and efficiency, yet effective assessment of SOH remains challenging for real-world applications (e.g., electric vehicle). In this paper, the estimation of SOH (i.e., capacity fading) under partial discharge with different starting and final State of Charge (SOC) levels is investigated. The challenge lies in the fact that partial discharge truncates the data available for SOH estimation, thereby leading to the loss or distortion of common SOH indicators. To address this challenge associated with partial discharge, we explore the convolutional neural network (CNN) to extract indicators for both SOH and changes in SOH  ($\Delta$SOH) between two successive charge/discharge cycles. The random forest algorithm is then adopted to produce the final SOH estimate by exploiting the indicators from the CNNs. Performance evaluation is conducted using the partial discharge data with different SOC ranges created from a fast-discharging dataset. The proposed approach is compared with i) a differential analysis-based approach and ii) two CNN-based approaches using only SOH and $\Delta$SOH indicators, respectively. Through comparison, the proposed approach demonstrates improved estimation accuracy and robustness. Sensitivity analysis of the CNN and random forest models further validates that the proposed approach makes better use of the available partial discharge data for SOH estimation.

\end{abstract}

%%%%%%%%%%%%%%%%%%%%%%%%%%%%%%%%%%%%%%%%%%%%%%%%%%%%%%%%%%%%%%%%%%%%%%%%%%%%%%%%%%%%
%%%%%%%%%%%%%%%%%%%%%%%%%%%%%%%%%%%%%%%%%%%%%%%%%%%%%%%%%%%%%%%%%%%%%%%%%%%%%%%%%%%%
\section{Introduction} \label{section.1}
The lithium-ion (Li-ion) battery has been well established as an effective energy storage technology for various applications due to its low self-discharge rate, high energy density, and falling cost~\cite{nykvist2015rapidly,song2018parameter}. To maintain safe and reliable operations, an accurate and robust battery State of Health (SOH) estimation is of critical importance. Generally, the battery SOH is characterized by the capacity or the internal resistance~\cite{han2019review}. Compared to resistance, capacity is a more direct indicator for SOH, as it represents the energy storage capability of a battery~\cite{li2019data}. Therefore, the estimation of capacity fading is the focus of most SOH monitoring works. To assess the SOH of a battery, one can completely discharge a fully charged battery and compute its capacity using coulomb counting~\cite{ng2009enhanced}. However, fully discharging or charging the battery for SOH estimation may not be feasible in some applications (e.g., electric vehicle) and can accelerate battery degradation~\cite{correa2015study}. Therefore, for general applications, SOH estimation is performed using the battery partial charge or discharge operating data. 

To estimate SOH using partial charge or discharge data, a mapping from the battery operating data (e.g., current, voltage, and temperature over time) to SOH needs to be constructed. However, the direct construction of this mapping requires a large amount of data, given the high dimensionality of this mapping. Without a comprehensive set of data, this mapping can be established with a two-step procedure. First, SOH indicators (e.g., parameters in physics-based models or signatures in experimental data) are identified from charge or discharge data based on expert knowledge. Then, a low-dimensional mapping is developed between these indicators and SOH~\cite{xiong2018towards}. 

In~\cite{chen2014application,wei2016adaptive,bi2016state}, equivalent circuit models (ECMs) were used with different filtering techniques to estimate open circuit voltage (OCV) versus State of Charge (SOC) curves. Then, SOH was inferred based on the relationship between OCV, SOC, and SOH. ECM-based approaches provide a computationally-efficient SOH estimation. However, given the complex electrochemical nature of batteries, the oversimplified ECMs will result in significant unmodeled dynamics, leading to degraded estimation accuracy. In addition, ECM-based approaches have fundamental difficulties extracting SOH indicators from the flat region of OCV-SOC curves, i.e., the mid-SOC range~\cite{song2018current}. 

To better capture the electrochemical mechanisms in batteries, electrochemical models were adopted for extracting SOH indicators~\cite{merla2018easy,uddin2016characterising}. Various physics-informed model parameters, such as solid electrolyte interphase growth and solid state diffusion coefficient, can be fitted from the battery operating data using the electrochemical models. SOH then is tracked based on its correlation with these model parameters~\cite{dey2014combined,li2018single}. The intensive computation of electrochemical-model-based approaches, however, makes them infeasible for most real-time applications~\cite{kim2011hybrid}. 

For a trade-off between computational complexity and model accuracy, differential analysis (DA), such as incremental capacity analysis (ICA) and differential voltage analysis, was utilized in~\cite{dubarry2009identify,weng2013board,han2014comparative}. Data-driven DA-based approach can extract SOH indicators better representing the true battery condition. However, due to the use of differential operations, data smoothing is required before applying DA, which could lead to distortion or loss of information. Moreover, SOH indicators from DA are mostly located near the high or low SOC range, thereby limiting its effectiveness under partial charge or discharge. 

Considering the aforementioned challenges in extracting SOH indicators and the increasing availability of data, deep-learning-based approaches offer a promising alternative to the above methods. By leveraging the strong learning ability of neural networks (NNs), deep-learning-based approaches can directly approximate the mapping from charge or discharge data to SOH (i.e., end-to-end estimation) without relying on expert knowledge~\cite{chaoui2017state,you2017diagnosis,shen2019deep}. However, one drawback for the deep-learning-based approaches that perform end-to-end estimation is the lack of interpretability, as there is no explicit SOH indicator extraction procedure. In addition, it is intrinsically difficult to incorporate expert knowledge into the estimation procedure for these approaches.

This paper proposes a deep-learning-based SOH estimation algorithm for a single battery cell under partial discharge. The proposed approach adopts a two-step procedure for SOH estimation to facilitate better interpretability and the incorporation of expert knowledge. In the first step, motivated by the work in~\cite{shen2019deep}, convolutional NN (CNN) is utilized to handle SOH indicator extraction under partial discharge. Two CNNs are established to extract the indicators related to the SOH and the change of SOH between two successive discharge cycles ($\Delta$SOH). In the second step, considering the potential correlation among indicators extracted from two CNNs, a random forest model is developed to produce the final SOH estimate based on the extracted indicators. To validate the effectiveness of the proposed approach, it is compared with the following approaches: i) a DA-based approach using ICA for SOH indicator extraction, and ii) two CNN-based approaches using SOH and $\Delta$SOH indicators, respectively, for SOH estimation. Based on a fast-discharging dataset provided in~\cite{severson2019data}, it has been demonstrated that the proposed approach can improve both estimation accuracy and robustness.

The contribution of this paper is three-fold. Firstly, the proposed SOH estimation algorithm based on CNN and random forest provides a general framework that can leverage both the learning ability and physics-based knowledge. Secondly, sensitivity analysis of CNN and random forest models is performed to interpret the proposed framework and show its effectiveness. Finally, a new perspective for SOH estimation that extracts $\Delta$SOH-related indicators is proposed, which could enhance the SOH detectability under partial charge or discharge conditions. 

The remainder of this paper is organized as follows: Section~\ref{section.2} introduces the procedure for partial discharge data collection and formulates the SOH estimation problem under partial discharge. The limitation of the DA-based approach under partial discharge is shown in Section~\ref{section.3}. Two CNNs extracting indicators related to SOH and $\Delta$SOH are developed, along with the examination of their performance, in Section~\ref{section.4}. In Section~\ref{section.5}, the sensitivity analysis of CNNs is performed which reveals the potential for combining two CNNs, and a random forest model is developed to fuse the outputs from the CNNs to enhance SOH estimation. In Section~\ref{section.6}, the proposed approach is verified under partial discharge with different initial and final SOC levels. Conclusions and future work are provided in Section~\ref{section.7}.

%%%%%%%%%%%%%%%%%%%%%%%%%%%%%%%%%%%%%%%%%%%%%%%%%%%%%%%%%%%%%%%%%%%%%%%%%%%%%%%%%%%%
%%%%%%%%%%%%%%%%%%%%%%%%%%%%%%%%%%%%%%%%%%%%%%%%%%%%%%%%%%%%%%%%%%%%%%%%%%%%%%%%%%%%
\section{Background} \label{section.2}
In this section, we first present the data to be used for developing and validating the proposed approach. Then, the SOH estimation problem under partial discharge is formulated. In this study, an open-source dataset provided in~\cite{severson2019data}, which contains aging data for 124 commercial batteries, is used to demonstrate the concept and validate the approach. The batteries under investigation are lithium-iron-phosphate/graphite cells manufactured by A123 Systems (model APR18650M1A). The nominal capacity of the cells is 1.1 $Ah$, and the lower and upper cut-off voltages are 2.0 $V$ and 3.6 $V$, respectively. The battery is cycled in a temperature-controlled environmental chamber of $\SI{30}{\celsius}$ by charging and discharging the cell repeatedly. A varied charging rate and a constant discharging rate are adopted for cycling. We consider using the discharge data for the subsequent study to focus on addressing the SOH estimation under incomplete charge and discharge conditions (instead of different C rates). 

\begin{figure}[!ht]
\centering
\includegraphics[width=3.3in]{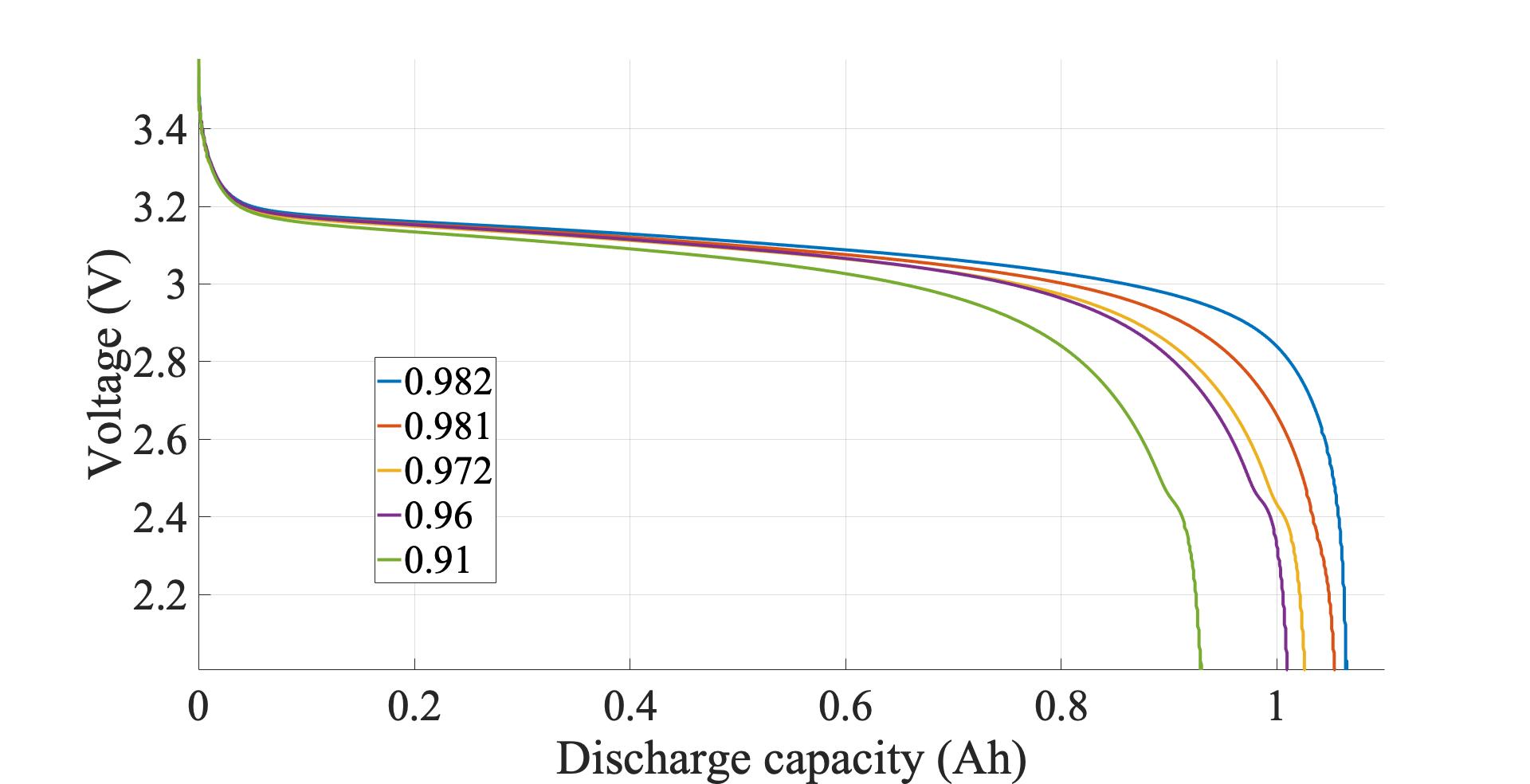}
\caption{Full discharge curves under different SOH levels} 
\label{fig:Discharging curves under different SOHs} 
\end{figure}

The cells are discharged from a fully charged state with a CC-CV policy, i.e., 4 $C$ to 2.0 $V$ with a current cutoff of 1/50 $C$. The aging data consists of the discharge capacity, voltage and cell capacity of each discharge cycle, where the cell capacity is obtained as the maximum discharge capacity over the entire cycle. Based on the cell capacity, the SOH for each discharge cycle is defined as the ratio between the cell capacity and the nominal capacity. The full discharge capacity/voltage curves (i.e., full discharge curves) from one sample cell under different SOH levels are given in Fig.~\ref{fig:Discharging curves under different SOHs}.

\begin{figure}[!ht]
\centering
\includegraphics[width=2.8in]{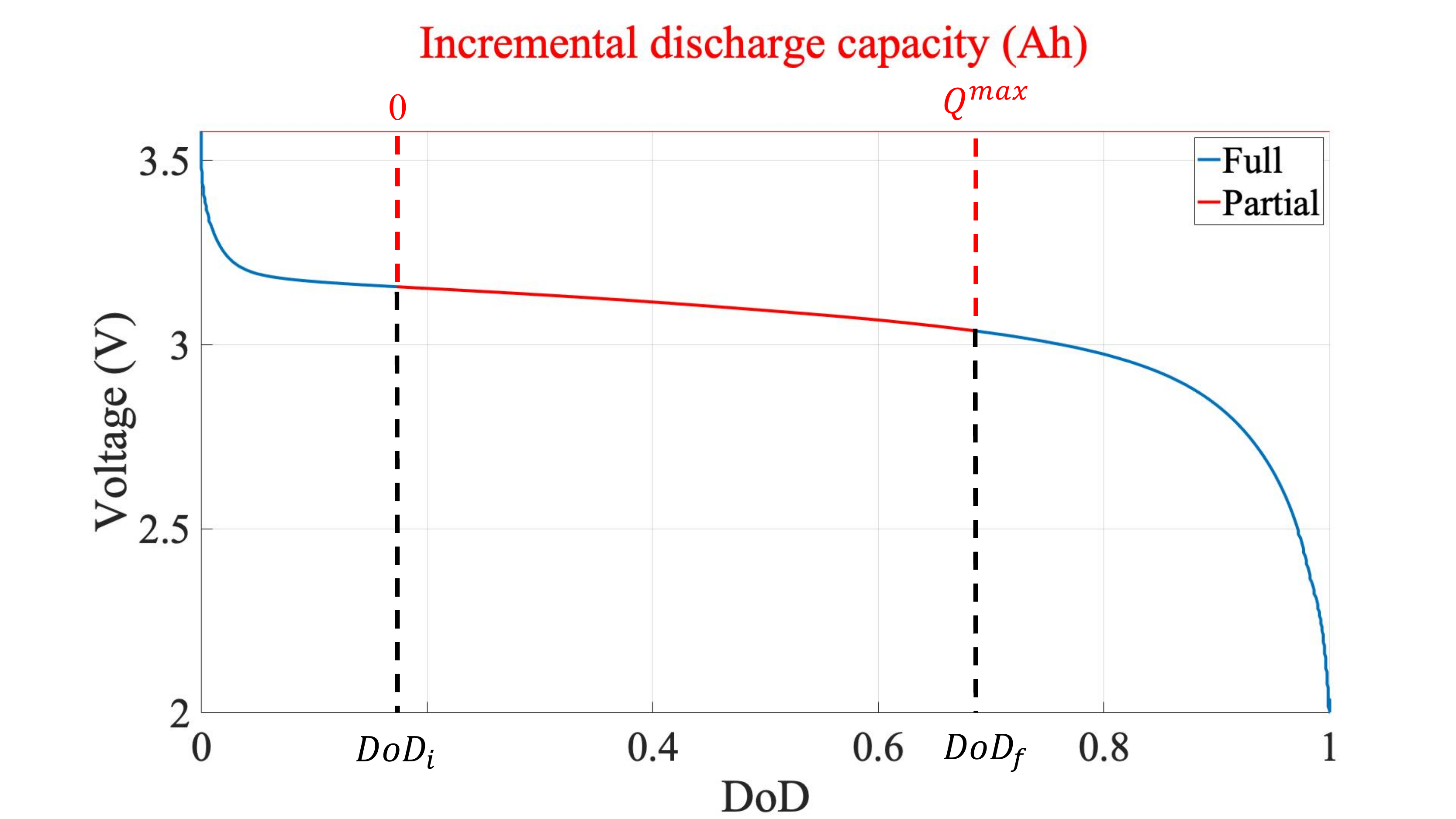}
\caption{Illustration of data truncation for creating partial discharge curve} 
\label{fig:Illustration of a partial discharge curve} 
\end{figure}

Using the full discharge curves, we create partial discharge dataset by truncating the full discharge curves with different initial and final depth of discharge (DoD) values. The initial DoD value ($DoD_{i}$) is determined randomly according to a Gaussian distribution, and the final DoD ($DoD_{f}$) is obtained as
\begin{equation} \label{eq:relationship initial and final DoD} 
DoD_{f} = DoD_{i} + \frac{Q^{max}}{C_{cell}},
\end{equation}
where $Q^{max}$ is a uniformly sampled maximum incremental discharge capacity, and $C_{cell}$ is the cell capacity. Note that, $Q^{max} = 0$ implies that $DoD_{f} = DoD_{i}$, namely, no discharge occurs in the present cycle. Then, the voltage (V) and incremental discharge capacity (Q) sequences from the initial to final DoD values are considered as the partial discharge curve for each cycle (see Fig.~\ref{fig:Illustration of a partial discharge curve} for an example). The SOH estimation problem under partial discharge is formulated as estimating the SOH for the present discharge cycle based on the present and past partial discharge curves.

%%%%%%%%%%%%%%%%%%%%%%%%%%%%%%%%%%%%%%%%%%%%%%%%%%%%%%%%%%%%%%%%%%%%%%%%%%%%%%%%%%%%
%%%%%%%%%%%%%%%%%%%%%%%%%%%%%%%%%%%%%%%%%%%%%%%%%%%%%%%%%%%%%%%%%%%%%%%%%%%%%%%%%%%%
\section{DA-based approaches for SOH estimation under partial discharge} \label{section.3}

In this section, we apply a DA-based approach to estimate SOH under partial discharge to illustrate the limitation of DA-based approaches and the characteristics of SOH indicators. The DA-based approach that uses ICA for indicator extraction is adopted given its demonstrated effectiveness for SOH estimation under partial charge/discharge~\cite{dubarry2006incremental,weng2016state,li2018quick,li2019state}. Typical ICA takes the following steps to extract SOH indicators from the discharge curves. First, data pre-processing is performed to smooth out the measurement noise. In this study, we utilize the support vector regression (SVR) method proposed in~\cite{weng2013board,weng2016state}, which fits the discharge curves (i.e., V-Q curves) for smoothing. The choice of SVR is based on the finding in~\cite{weng2013board,weng2016state} that SVR can effectively remove noise and has minimal information loss under partial charge/discharge. With the smoothed discharge curves, the IC values are then computed as the gradient of V with respect to Q (i.e., $\Delta$Q/$\Delta$V). Finally, by examining the IC/voltage curves, salient features can be extracted as the SOH indicators. 

\begin{figure}[!ht]
\centering
\includegraphics[width=3.3in]{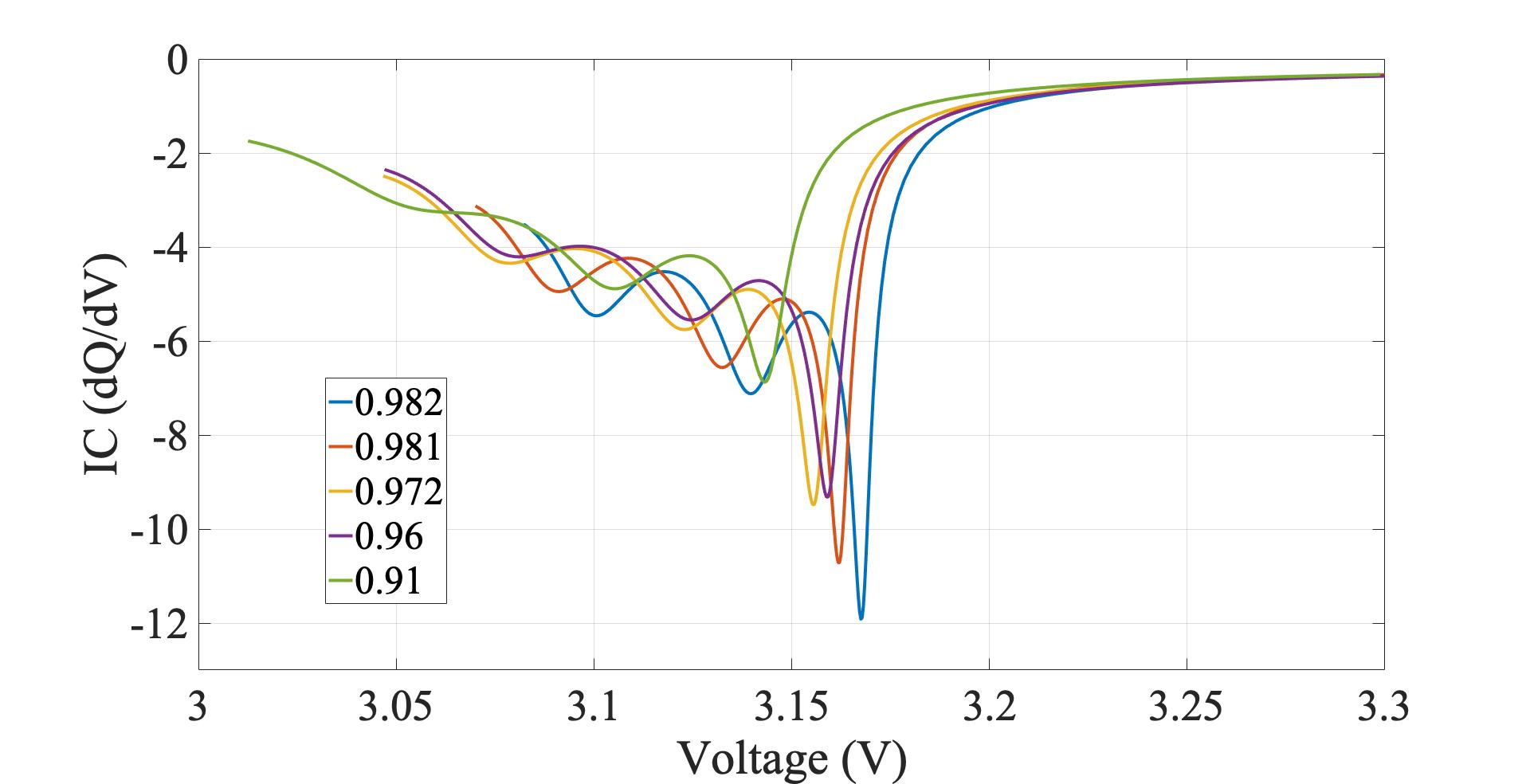} 
\caption{IC curves under different SOH levels for the cell in Fig.~\ref{fig:Discharging curves under different SOHs} under partial discharge with the low DoD region} 
\label{fig:IC curves FS1} 
\end{figure}

Considering that the IC features are mostly in the high SOC (i.e., low DoD) region~\cite{weng2013board}, we create a partial discharge dataset with $DoD_{i} = 0$ and $Q^{max} \in \mathcal{U}(0.65,0.75)$, where $\mathcal{U}$ denotes a uniform distribution. By applying SVR and differentiation to the partial discharge curves, the IC curves for the same cell in Fig.~\ref{fig:Discharging curves under different SOHs} are shown in Fig.~\ref{fig:IC curves FS1}. From Fig.~\ref{fig:IC curves FS1}, we have the following observations:
\begin{itemize}
    \item There are multiple IC minima over the available voltage range, which correlate with SOH.
    \item The location of these IC minima shifts irregularly along the voltage axis under different SOH levels.
    \item Only the lowest IC minimum remains identifiable under different SOH levels.
\end{itemize}
The above observations indicate that, under partial discharge, the SOH estimation performance of the DA-based approach that uses ICA heavily relies on the existence and the quality of the lowest IC minimum. 

\begin{figure}[!ht]
\centering
\includegraphics[width=3.3in]{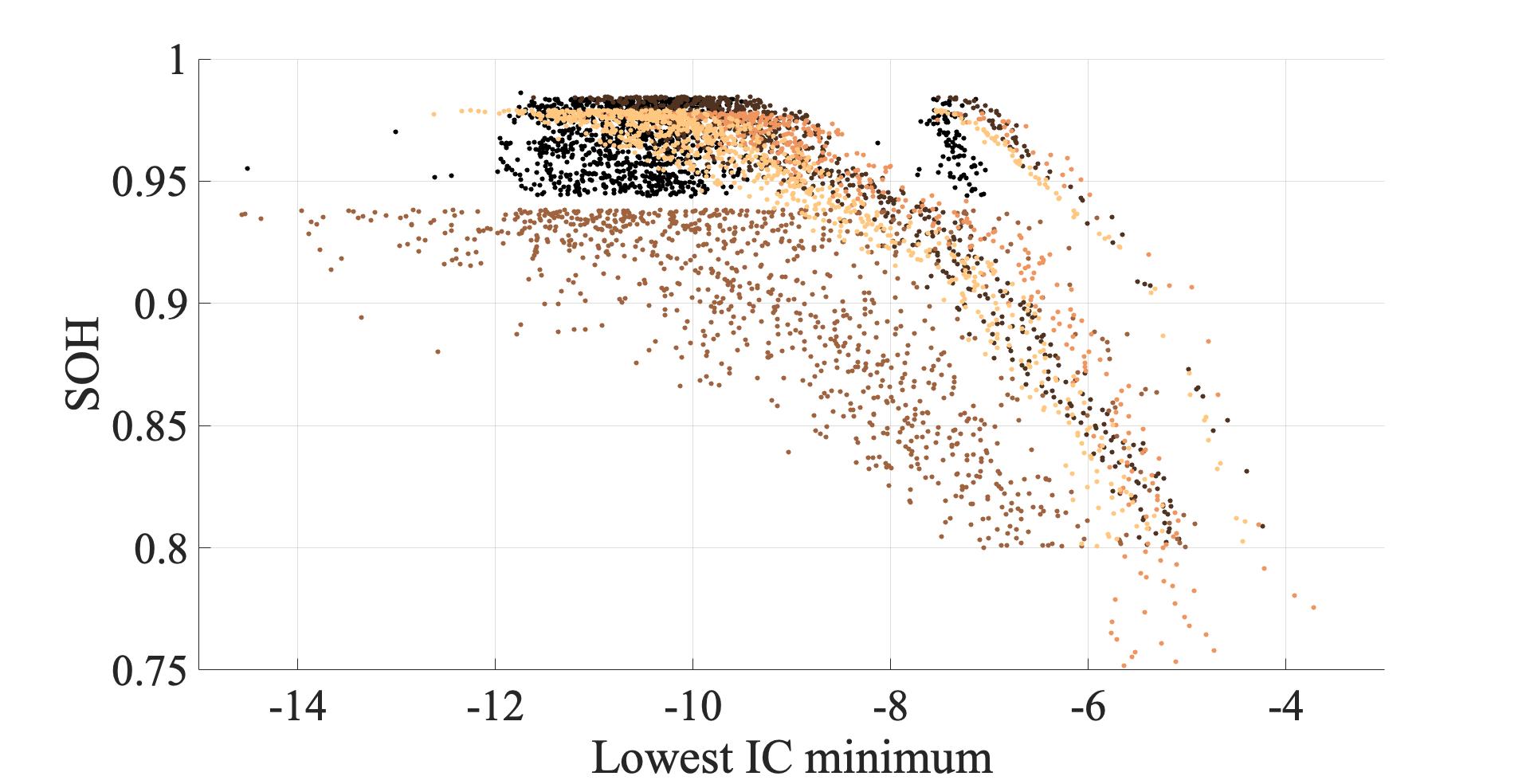}
\caption{The largest IC peak values vs. SOH from five different cells under partial discharge with the low DoD region} 
\label{fig:IC peaks FS1} 
\end{figure}

To examine the quality of the lowest IC minimum for different cells, we plot the lowest IC minimum for five different cells and its corresponding SOH levels in Fig.~\ref{fig:IC peaks FS1}, where the dots with the same color indicate the data from the same cell. It can be seen from Fig.~\ref{fig:IC peaks FS1} that: 
\begin{itemize}
    \item The lowest IC minimum of the cell may not be consistently correlated with its SOH.
    \item The relationship between the lowest IC minimum and SOH can vary for different cells.
\end{itemize}
Since the correlation between the lowest IC minimum and SOH is corrupted under partial discharge and sensitive to the cell-to-cell variation, the DA-based approach can suffer from performance degradation when estimating the SOH. 

\begin{figure}[!ht]
\centering
\includegraphics[width=3.3in]{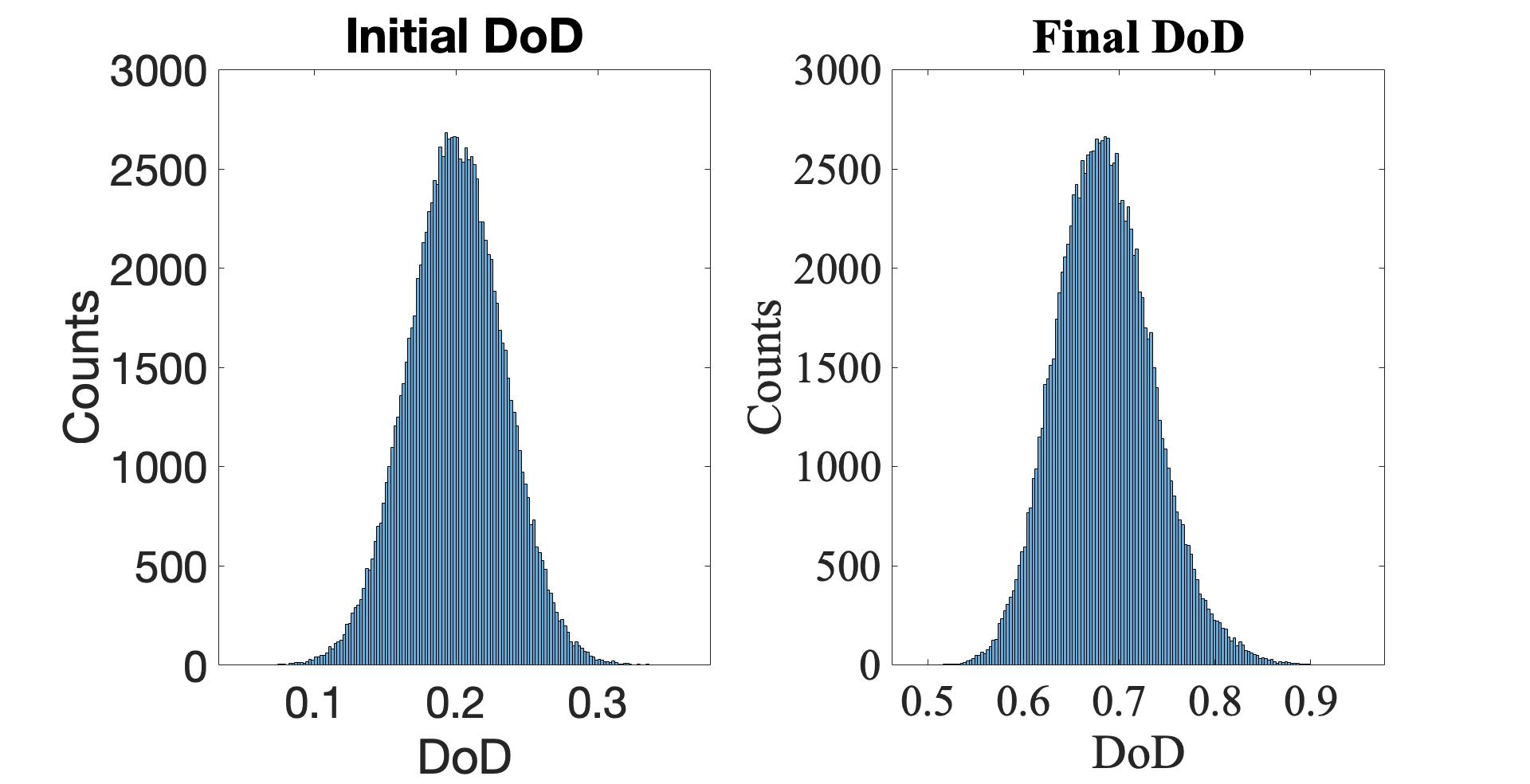}
\caption{Initial DoD and final DoD distributions for the partial discharge dataset without the low DoD region} 
\label{fig:FS2 DoD Distributon} 
\end{figure}

\begin{figure}[!ht]
\centering
\includegraphics[width=3.3in]{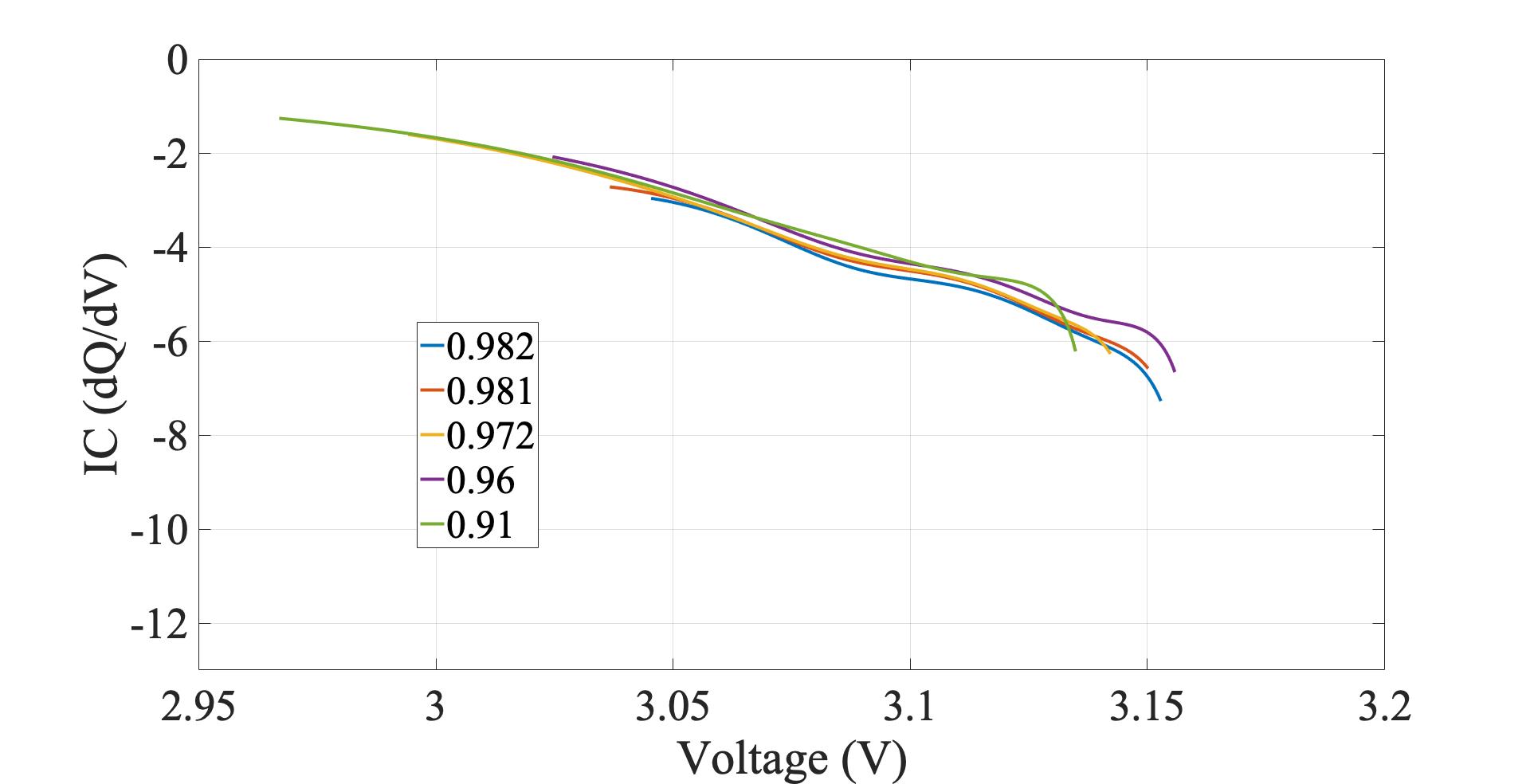} 
\caption{IC curves under different SOH levels for the cell in Fig.~\ref{fig:Discharging curves under different SOHs} under partial discharge without the low DoD region} 
\label{fig:IC curves FS2} 
\end{figure}

To further illustrate the robustness issue with the DA-based approach under partial discharge, we create another partial discharge dataset with $DoD_{i} \in \mathcal{N}(0.2,1/900)$ and $Q^{max} \in \mathcal{U}(0.45,0.55)$, respectively, where $\mathcal{N}$ denotes a Gaussian distribution. The distributions for the sampled initial and final DoD values are shown in Fig.~\ref{fig:FS2 DoD Distributon}. This partial discharge setup represents a normal battery operating condition in a typical electric vehicle application. The IC curves from this dataset for the same cell in Fig.~\ref{fig:IC curves FS1} is plotted in Fig.~\ref{fig:IC curves FS2}. In Fig.~\ref{fig:IC curves FS2}, we see that:
\begin{itemize}
    \item Due to the data truncation caused by partial discharge, the lowest IC minimum between 3.15 $V$ and 3.2 $V$ disappears. 
    \item The IC minima located between 3.05 $V$ and 3.15 $V$ in Fig.~\ref{fig:IC curves FS1} also disappear due to the data smoothing.
\end{itemize}

%%%%%%%%%%%%%%%%%%%%%%%%%%%%%%%%%%%%%%%%%%%%%%%%%%%%%%%%%%%%%%%%%%%%%%%%%%%%%%%%%%%%
%%%%%%%%%%%%%%%%%%%%%%%%%%%%%%%%%%%%%%%%%%%%%%%%%%%%%%%%%%%%%%%%%%%%%%%%%%%%%%%%%%%%
\section{CNNs for SOH estimation under partial discharge} \label{section.4}
In this section, CNNs are developed to address SOH estimation under partial discharge. First, we develop a CNN that directly estimates SOH from the discharge data and evaluate its performance under partial discharge without the low DoD region. In view of the low consistency between successive SOH estimates from the CNN for direct SOH estimation, a CNN for incremental SOH estimation is then proposed.   

%%%%%%%%%%%%%%%%%%%%%%%%%%%%%%%%%%%%%%%%%%%%%%%%%%%%%%%%%%%%%%%%%%%%%%%%%%%%%%%%%%%%
\subsection{Direct SOH estimation using CNN} \label{section.4A}
In Section~\ref{section.3}, it has been shown that DA-based approaches become ineffective when the expert knowledge (e.g., IC minima) is not applicable. In the absence of applicable expert knowledge, NN can be a promising solution, since it can automatically extract features from the data that correlate with the desired output. For SOH estimation, three types of NNs have been investigated, i.e., multilayer perception (MLP)~\cite{andre2013comparative}, recurrent NN (RNN)~\cite{chaoui2017state,you2017diagnosis}, and CNN~\cite{shen2019deep}. In order to select the best NN for our problem, we recall the observations in Section~\ref{section.3} that the IC minimum is a local property and may shift when the SOC range of the partial discharge curves differs. Consequently, CNN is chosen to extract locally-connected features whose locations in the input data may shift~\cite{zhang1990parallel}. The advantage of CNN over MLP and RNN lies in its convolution and max-pooling operations. The former enables the extraction of locally-connected features such as the regional minima in IC curves, and the latter can identify shifted features by taking the extreme value within a local region as its output.

\begin{figure}[!ht]
\centering
\includegraphics[width=3.0in]{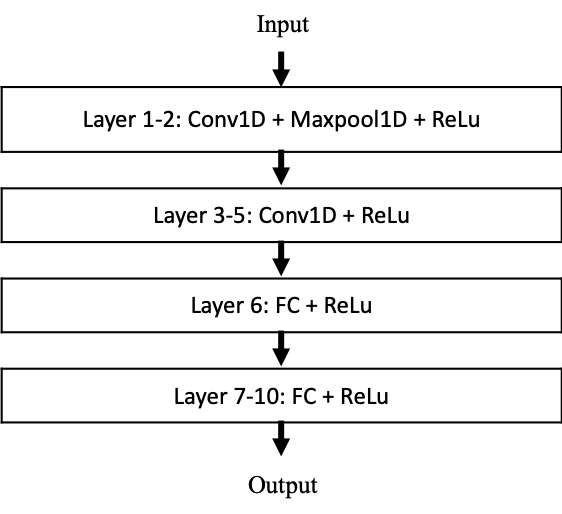}
\caption{Architecture for CNN} 
\label{fig:Architecture for CNN} 
\end{figure}

We use a ten-layer CNN model with both convolutional and fully-connected (FC) layers (see Fig.~\ref{fig:Architecture for CNN}). The Conv1D and Maxpool1D denotes the 1D convolution and max pooling operations. The hyperparameters for the CNN are listed in Table~\ref{table:Hyperparameters for CNN}. The input and output of the CNN are the partial discharge curve and SOH for each discharge cycle. To train the CNN, we use $60\%$ of the data for training, $20\%$ for validation, and $20\%$ for testing. Five-folds cross-validation is performed to remove the effect of the dataset partition on the estimation accuracy. For performance evaluation, we adopt the following definition of mean absolute error (MAE) between the true SOH ($SOH_t$) and the estimated SOH ($SOH_e$):
\begin{equation} \label{eq:SOH MAE definition} 
MAE = \frac{|SOH_t-SOH_e|}{SOH_t} \times 100 \%.
\end{equation}
\begin{table}
\begin{spacing}{1.0}
\centering 
\caption{Hyperparameters for CNNs}
\label{table:Hyperparameters for CNN}
\begin{tabular}{@{}cc@{}} 
\toprule
Hyperparameter & Value  \\ \midrule
Convolution filter size &  3  \\
Convolution filter stride  &   1   \\ 
Number of convolution filters  &   50   \\ 
Pooling size  &    3    \\
Pooling stride     &   3  \\ 
Layer 6 FC dimension & 550 \\
Layer 7-10 FC dimension & 200 \\ \bottomrule
\end{tabular} 
\end{spacing} 
\end{table} 

\begin{figure*}
  \includegraphics[width=\textwidth]{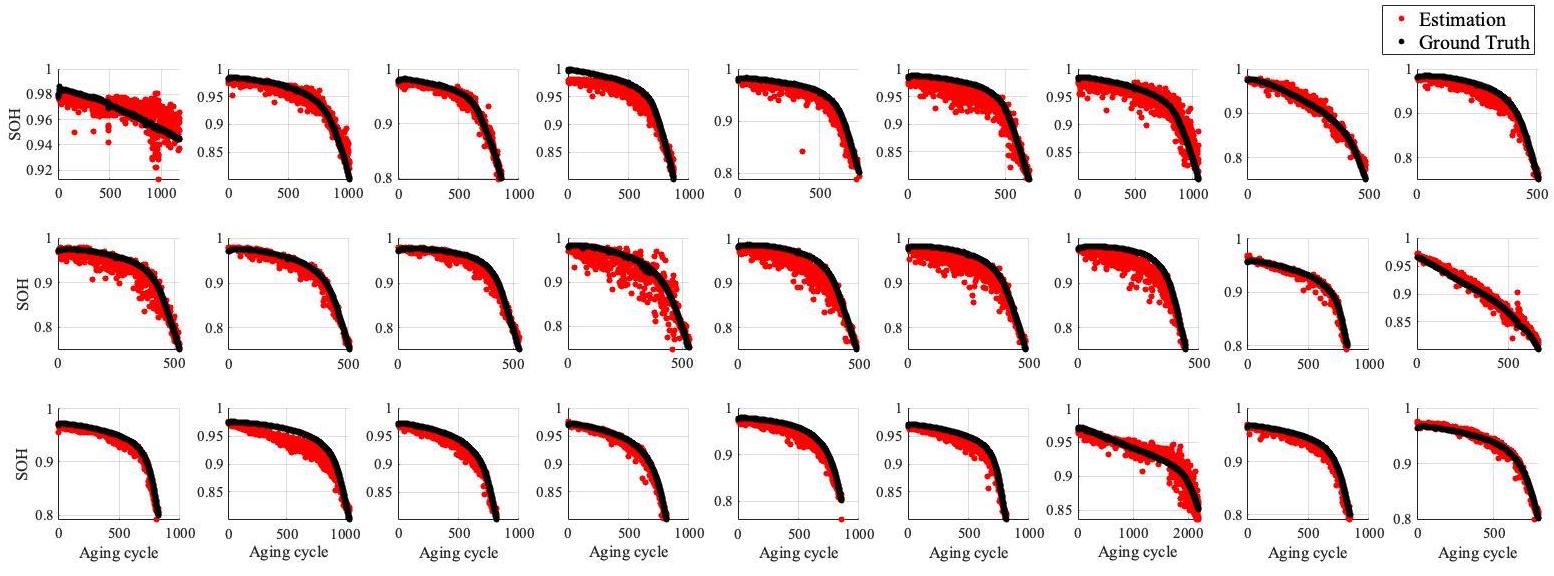}
  \caption{Representative SOH estimation performance from SOH-CNN under partial discharge without the low DoD region (Each subplot contains the SOH trajectory of one cell)}
  \label{fig:Representative performance SOH FS2} 
\end{figure*}

We apply the CNN with the above setup (SOH-CNN) to the same dataset used for DA-based approach, namely, with the initial and final DoD values located around 0.2 and 0.7 (see Fig.~\ref{fig:FS2 DoD Distributon}). The average MAE from cross-validation using SOH-CNN is $1.28\%$, and its representative SOH estimation performance is presented in Fig.~\ref{fig:Representative performance SOH FS2}. As is shown in Fig.~\ref{fig:Representative performance SOH FS2}, a satisfactory SOH estimation performance can be achieved by SOH-CNN without explicitly incorporating the salient SOH indicator used in the DA-based approach (i.e., the lowest IC minimum). This is because the indicators that are numerically correlated with SOH, but may not be physically interpretable, can be extracted by SOH-CNN. In addition, since no data smoothing is required, the potential information loss caused by data processing can be avoided. Despite its success in handling SOH estimation under partial discharge, it does not yield consistent SOH estimates in successive cycles, as shown in Fig.~\ref{fig:Representative performance SOH FS2}. This is due to the following two reasons. First, the information provided with the partial discharge curve may not be enough, thereby leading to a degraded estimation accuracy. Second, only the partial discharge data of present cycle is used for SOH estimation. Consequently, the correlation between present and past SOH values can not be captured.

%%%%%%%%%%%%%%%%%%%%%%%%%%%%%%%%%%%%%%%%%%%%%%%%%%%%%%%%%%%%%%%%%%%%%%%%%%%%%%%%%%%%
\subsection{Incremental SOH estimation using CNN} \label{section.4B}
To enhance consistency among successive SOH estimates, we seek a CNN formulation that can better capture the accumulative and evolutionary nature of SOH. To this end, we propose to use the difference of SOH between the present and past SOHs ($\Delta$SOH) as the CNN output, in an attempt to extract indicators related to the change in SOH. Since the change of SOH depends on the degradation history, the past partial discharge curve is used with the present discharge curve as additional input channels to CNN. The same CNN architecture and training procedure discussed in Section~\ref{section.4A} are adopted. Note that we do not constrain the $\Delta$SOH from the CNN to be strictly negative for a monotonically decreasing SOH trajectory, as $\Delta$SOH can be positive due to various factors~\cite{severson2019data}. 
%In particular, for the fast-discharging dataset used in this study, since the SOH is measured for each discharge cycle, the $\Delta$SOH can be both positive and negative (e.g., see Fig.~\ref{fig:dSOH example})
%
% \begin{figure}[!ht]
% \centering
% \includegraphics[width=3.3in]{Plots/dSOH_example.jpg}
% \caption{$\Delta$SOH history for the cell given in Fig.~\ref{fig:Discharging curves under different SOHs}} 
% \label{fig:dSOH example} 
% \end{figure}
Based on the estimated $\Delta$SOH from CNN, the SOH estimate for the present cycle ($SOH^{t}_e$) can be computed as the sum of the SOH estimate for the past cycle ($SOH^{t-1}_e$) and the estimated $\Delta$SOH for the present cycle ($\Delta SOH_e^{t}$), i.e.,
\begin{equation} \label{eq:SOH estimation in dSOH formulation} 
SOH^{t}_e = SOH^{t-1}_e + \Delta SOH_e^{t}.
\end{equation}

\begin{figure*}
  \includegraphics[width=\textwidth]{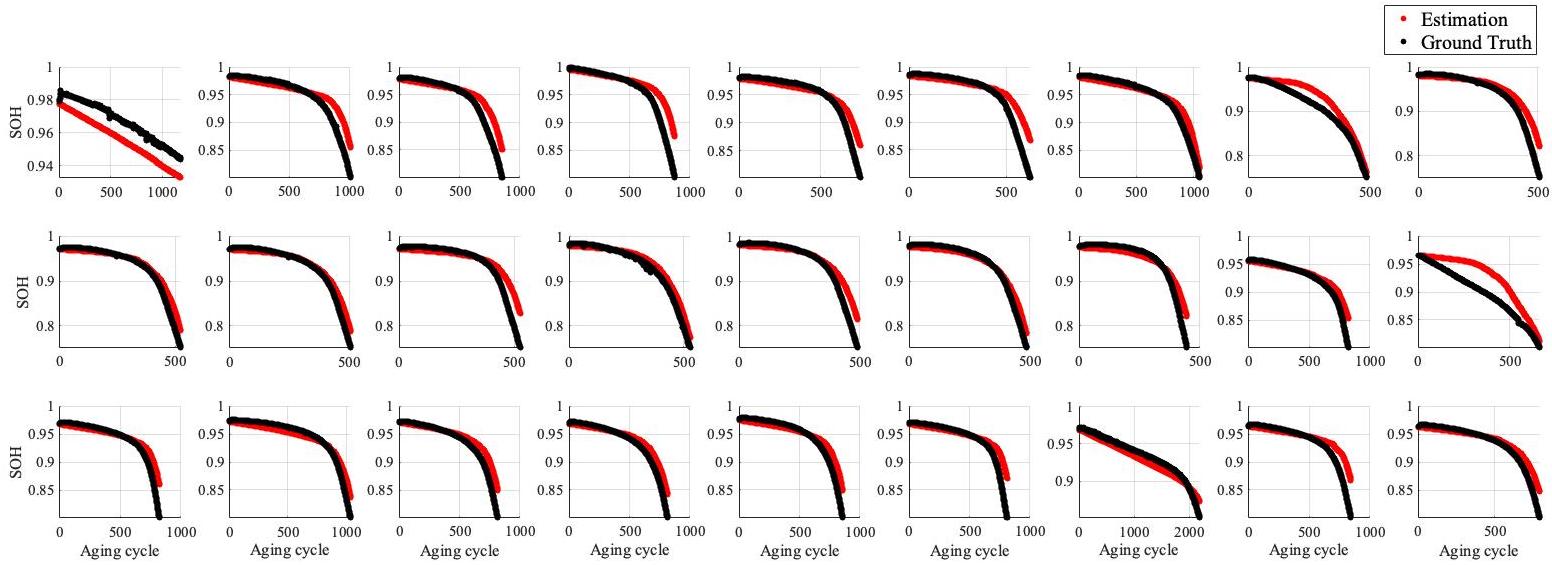}
  \caption{Representative SOH estimation performance from $\Delta$SOH-CNN under partial discharge without the low DoD region (Each subplot contains the SOH trajectory of one cell)}
  \label{fig:Representative performance dSOH FS2} 
\end{figure*}

For the same dataset used in Section~\ref{section.4A}, the average MAE from the CNN that outputs $\Delta$SOH ($\Delta$SOH-CNN) is $1.57\%$, and one representative SOH estimation result is shown in Fig.~\ref{fig:Representative performance dSOH FS2}. Compared to the SOH estimation result in Fig.~\ref{fig:Representative performance SOH FS2}, $\Delta$SOH-CNN improves the consistency among the successive SOH estimates. This is attributed to two reasons. First, as both the present and past discharge curves are used as inputs to $\Delta$SOH-CNN, richer information is available for estimating SOH. More importantly, by formulating the output of the CNN as $\Delta$SOH, $\Delta$SOH-CNN could better learn the relationship between successive SOHs by extracting $\Delta$SOH-related features, leading to an improved consistency. On the downside, $\Delta$SOH-CNN has a worse MAE than that of SOH-CNN, as the estimation errors will be accumulated over time like any integrator-based estimators. Furthermore, since the $\Delta$SOH is typically much smaller than SOH, the ground truth $\Delta$SOH used for CNN training is more likely to be corrupted by noise in the experimental data, which could affect the training and deteriorate the estimation accuracy.

%%%%%%%%%%%%%%%%%%%%%%%%%%%%%%%%%%%%%%%%%%%%%%%%%%%%%%%%%%%%%%%%%%%%%%%%%%%%%%%%%%%% 
%%%%%%%%%%%%%%%%%%%%%%%%%%%%%%%%%%%%%%%%%%%%%%%%%%%%%%%%%%%%%%%%%%%%%%%%%%%%%%%%%%%%
\section{CNN and random forest for robust SOH estimation under partial discharge} \label{section.5}

In this section, we first perform sensitivity analysis of the two CNNs developed in Section~\ref{section.4} to understand the contribution of different portions in the discharge curve to the SOH estimation. Based on the CNN sensitivity analysis, a random forest model is then designed to integrate both CNNs for an enhanced SOH estimation performance. 

%%%%%%%%%%%%%%%%%%%%%%%%%%%%%%%%%%%%%%%%%%%%%%%%%%%%%%%%%%%%%%%%%%%%%%%%%%%%%%%%%%%%
\subsection{CNN model sensitivity analysis} \label{section.5A}
\begin{figure}[!ht]
\centering
\includegraphics[width=3.3in]{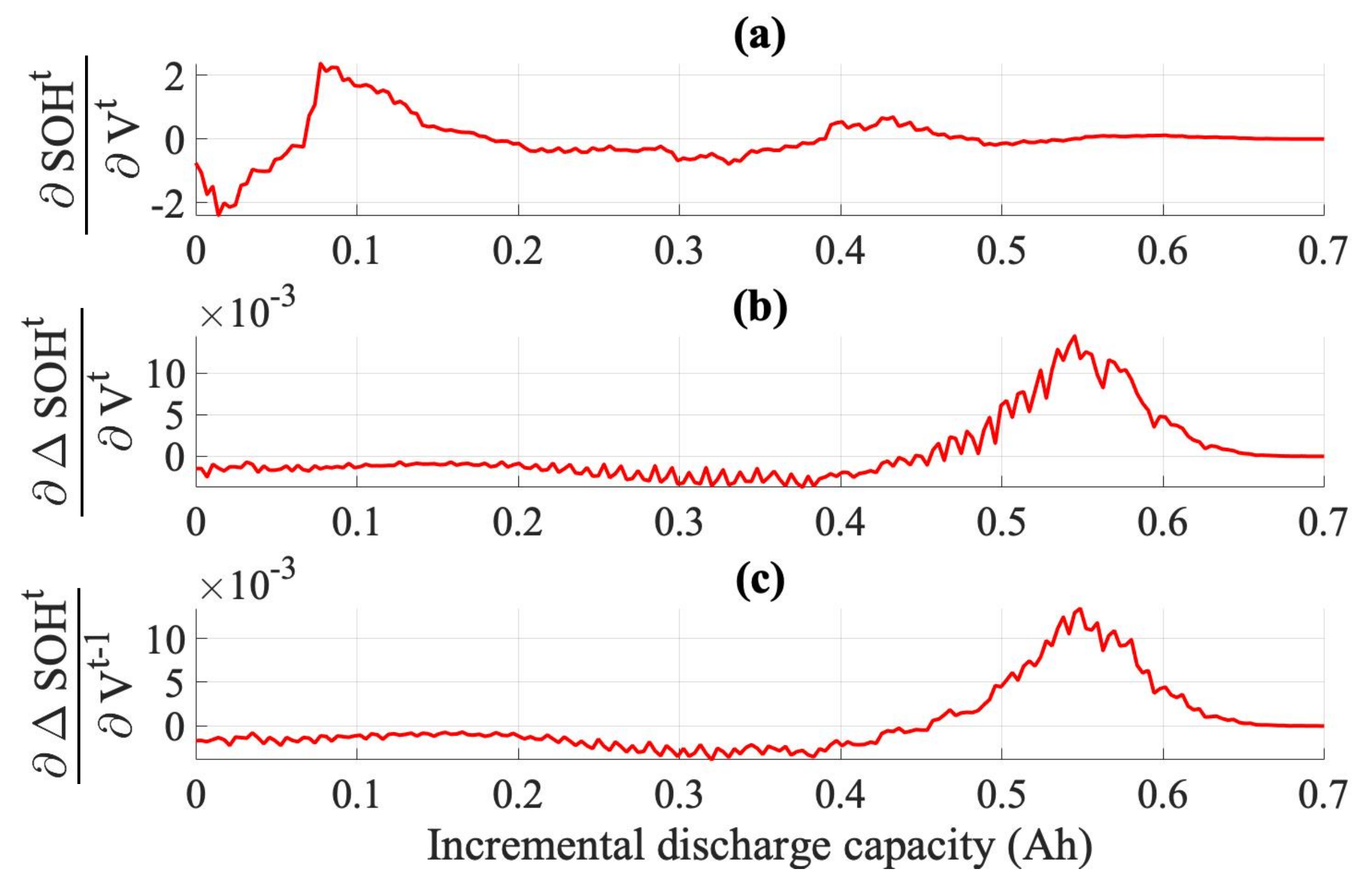}
\caption{Sensitivity of the output with respect to the input voltage curves in CNNs ($V_t$ and $V_{t-1}$ are the present and past voltage curves, (a) is for SOH-CNN, and (b) and (c) are for $\Delta$SOH-CNN)} 
\label{fig:CNN sensitivity} 
\end{figure}

As is shown in Section~\ref{section.4}, both SOH-CNN and $\Delta$SOH-CNN can handle SOH estimation under partial discharge and outperform the DA-based approach. The SOH-CNN has smaller MAE but worse consistency in terms of successive SOH estimates compared to $\Delta$SOH-CNN. Motivated by their complementary characteristics, we investigate if these two CNN models can supplement each other for SOH estimation. To this end, we analyze the CNN model sensitivity by computing the partial derivative of the output with respect to the inputs. In particular, considering that SOH is closely related to OCV, we study the sensitivity of the CNN output with respect to the input voltage curves, as plotted in Fig.~\ref{fig:CNN sensitivity}. It shows that SOH-CNN has larger partial derivative values in the first half of the present voltage curve under partial discharge (see Fig.~\ref{fig:CNN sensitivity}(a)), while $\Delta$SOH-CNN shows larger partial derivative values in the second half of the present and past voltage curves (see Fig.~\ref{fig:CNN sensitivity}(b) and Fig.~\ref{fig:CNN sensitivity}(c)). This observation motivates the random forest approach for combing these two CNN models presented in the next section. 

%%%%%%%%%%%%%%%%%%%%%%%%%%%%%%%%%%%%%%%%%%%%%%%%%%%%%%%%%%%%%%%%%%%%%%%%%%%%%%%%%%%%
\subsection{CNN and random forest for SOH estimation} \label{section.5B}
\begin{figure}[!ht]
\centering
\includegraphics[width=3.3in]{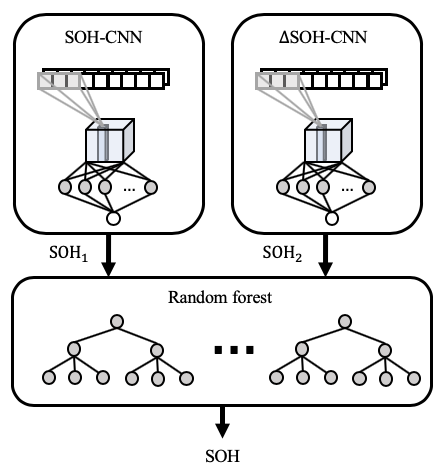}
\caption{Overall structure for the RF-CNN (SOH$_1$ and SOH$_2$ are the estimated SOH from SOH-CNN and $\Delta$SOH-CNN)} 
\label{fig:RF CNN structure} 
\end{figure}

To combine two CNN models for estimating SOH, we treat the SOH estimates from two CNNs as two SOH indicators. A regression model can then be developed to estimate SOH. Considering that the SOH estimates from CNNs are both produced based on the discharge curves, there can be strong correlation between these two SOH estimates. Therefore, the random forest algorithm is chosen to mitigate the potential effect caused by multicollinearity~\cite{farrar1967multicollinearity}. By combining the CNN and random forest models, the overall structure of our random forest-CNN (RF-CNN) SOH estimator is provided in Fig.~\ref{fig:RF CNN structure}. The development procedure starts by training the SOH-CNN and $\Delta$SOH-CNN. Then, a random forest model is trained to capture the optimal mapping from the outputs of both CNNs to the final estimate of SOH. In this study, the development of CNNs is performed in PyTorch~\cite{adam2017automatic} with Adam optimizer, and the development of the random forest is achieved using the statistics and machine learning toolbox in MATLAB.

\begin{figure*}
  \includegraphics[width=\textwidth]{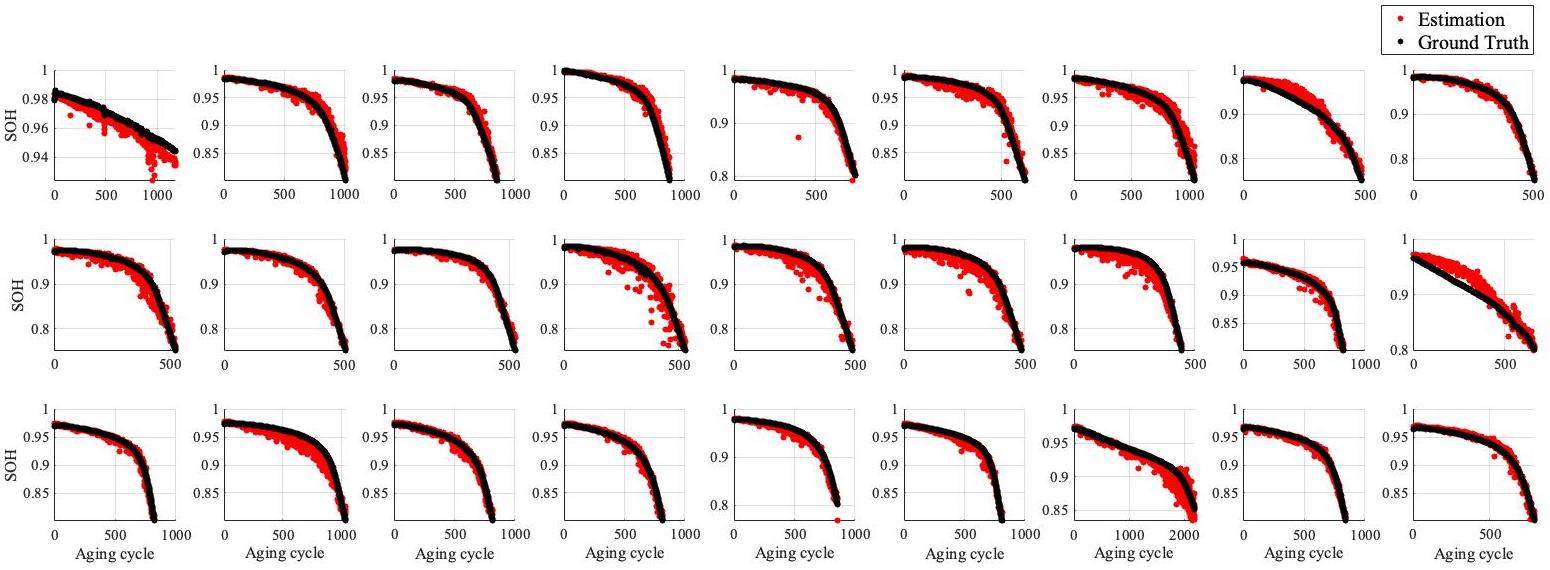}
  \caption{Representative SOH estimation performance from the RF-CNN under partial discharge without the low DoD region (Each plot contains the SOH trajectory for one cell)}
  \label{fig:Representative performance RF-CNN FS2} 
\end{figure*}

When compared to previous methods, the average MAE from RF-CNN is $0.85\%$ for the same case study in Section~\ref{section.4}, and its representative SOH estimation performance is shown in Fig.~\ref{fig:Representative performance RF-CNN FS2}. This shows that RF-CNN achieves a better consistency than SOH-CNN when compared to the result shown in Fig.~\ref{fig:Representative performance SOH FS2}. In addition, MAE is reduced by up to $35\%$ by RF-CNN when compared to SOH-CNN and $\Delta$SOH-CNN. To further illustrate the reason why RF-CNN can enhance the estimation performance, we study the importance of the outputs from two CNNs on the SOH estimate from the random forest model. In particular, the ratios of the importance of the SOH-CNN output to that of the $\Delta$SOH-CNN output over five folds in cross-validation are reported, which are 1.41, 1.38, 1.87, 1.53 and 1.00. From the reported ratios, it can be seen that both CNN models contribute substantially to the final SOH estimate, thereby resulting in a better estimation accuracy than that from each individual CNN. 
%In addition, the SOH-CNN contributes more in the random forest model, which might be caused by its lower MAE in SOH estimation. 
%

%%%%%%%%%%%%%%%%%%%%%%%%%%%%%%%%%%%%%%%%%%%%%%%%%%%%%%%%%%%%%%%%%%%%%%%%%%%%%%%%%%%% 
%%%%%%%%%%%%%%%%%%%%%%%%%%%%%%%%%%%%%%%%%%%%%%%%%%%%%%%%%%%%%%%%%%%%%%%%%%%%%%%%%%%%
\section{Performance evaluation} \label{section.6}
In this section, to validate the effectiveness of the proposed RF-CNN, we first compare it with the DA-based approach discussed in Section~\ref{section.3}. Then, an extensive comparison between RF-CNN, SOH-CNN and $\Delta$SOH-CNN is performed under partial discharge with different DoD ranges.

%%%%%%%%%%%%%%%%%%%%%%%%%%%%%%%%%%%%%%%%%%%%%%%%%%%%%%%%%%%%%%%%%%%%%%%%%%%%%%%%%%%%
\subsection{Comparison with DA based approaches} \label{section.6A}
\begin{figure*}
  \includegraphics[width=\textwidth]{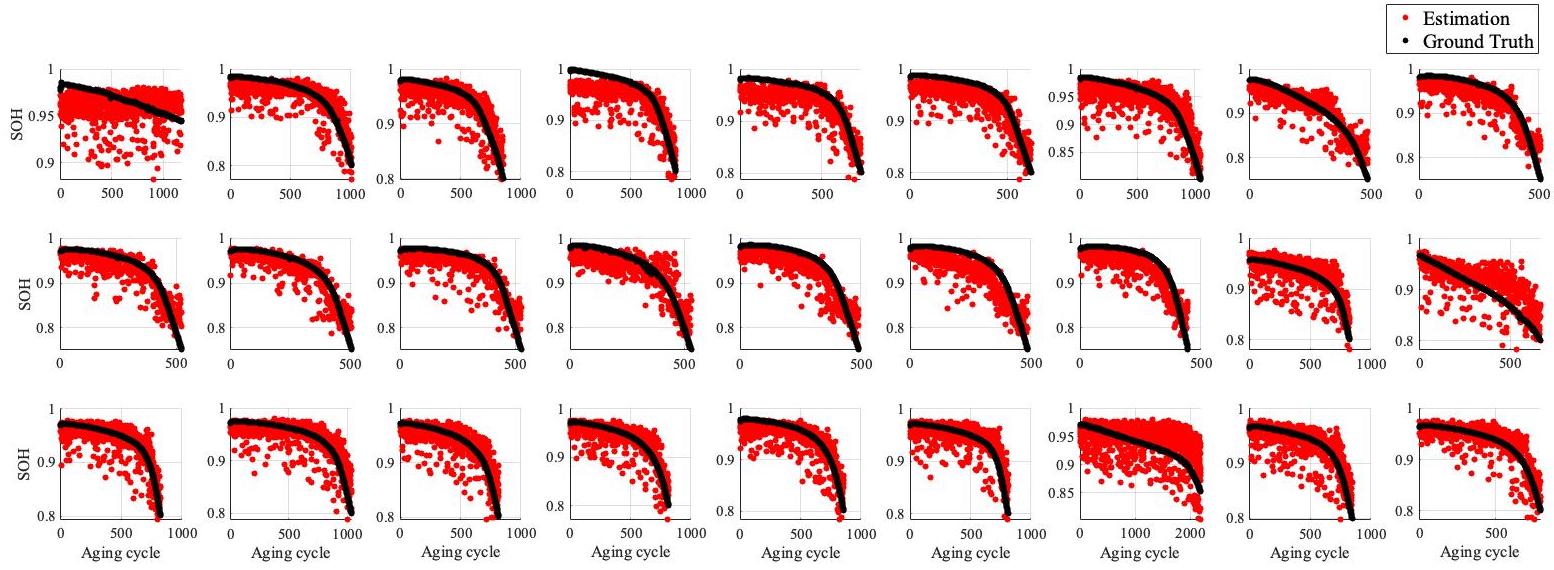}
  \caption{Representative SOH estimation performance from RF-ICA under partial discharge with the low DoD region (Each subplot contains the SOH trajectory of one cell)}
  \label{fig:Representative performance RF-ICA FS1} 
\end{figure*}

\begin{figure*}
  \includegraphics[width=\textwidth]{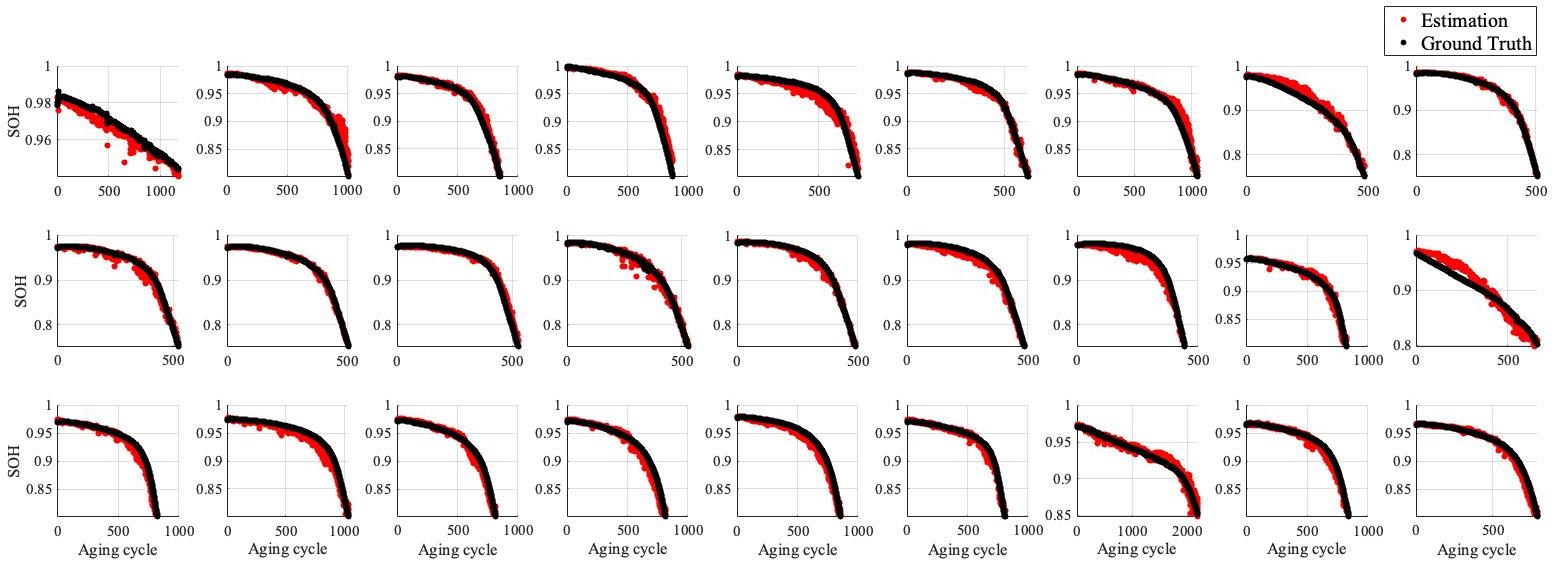}
  \caption{Representative SOH estimation performance from RF-CNN under partial discharge with the low DoD region (Each subplot contains the SOH trajectory of one cell)}
  \label{fig:Representative performance RF-CNN FS1} 
\end{figure*}

To perform comparison with the DA-based approach, we consider the partial discharge data with the low DoD region presented in Section~\ref{section.3}, i.e., $DoD_{i} = 0$. For a fair comparison, the random forest algorithm is adopted to establish the mapping from the lowest IC minimum to SOH. By applying the DA-based approach that uses ICA and random forest (RF-ICA), its average MAE is $2.31\%$, while the MAE from RF-CNN is $0.72\%$. Therefore, it is verified that RF-CNN can achieve better MAE under partial discharge thanks to the richness and robustness of the SOH indicators extracted by CNN. The representative results from RF-ICA and RF-CNN are shown in Fig.~\ref{fig:Representative performance RF-ICA FS1} and Fig.~\ref{fig:Representative performance RF-CNN FS1}, respectively. It shows that RF-ICA has worse consistency in successive SOH estimates because RF-ICA estimates SOH only based on the present discharge curve and therefore fails to capture the relationship between past and present SOHs. Furthermore, as is shown in Section~\ref{section.3} and \ref{section.5}, RF-CNN has better robustness compared to the RF-ICA under partial discharge without the low DoD region.

%%%%%%%%%%%%%%%%%%%%%%%%%%%%%%%%%%%%%%%%%%%%%%%%%%%%%%%%%%%%%%%%%%%%%%%%%%%%%%%%%%%%
\subsection{Comparison with SOH-CNN and $\Delta$SOH-CNN} \label{section.6B}
We further compare the RF-CNN with SOH-CNN and $\Delta$SOH-CNN under partial discharge with different DoD ranges to demonstrate the robustness of the RF-CNN and validate the benefit of combining the CNN models. The following four partial discharge conditions are considered: 
\begin{enumerate}[label=\roman*)]
    \item $DoD_{i} \in \mathcal{N}(0.1,1/900)$ and $Q^{max} \in \mathcal{U}(0.67,0.77)$; 
    \item $DoD_{i} \in \mathcal{N}(0.2,1/900)$ and $Q^{max} \in \mathcal{U}(0.45,0.55)$; 
    \item $DoD_{i} \in \mathcal{N}(0.3,1/900)$ and $Q^{max} \in \mathcal{U}(0.25,0.35)$; 
    \item $DoD_{i} \in \mathcal{N}(0.4,1/900)$ and $Q^{max} \in \mathcal{U}(0.05,0.15)$. 
\end{enumerate}
The resulted final DoD distribution for each condition will approximately be a Gaussian distribution as that in Fig.~\ref{fig:FS2 DoD Distributon} with its mean located near 0.8, 0.7, 0.6, and 0.5. By training and testing the estimators on each partial discharge dataset, the average MAE from different CNN-based estimators is summarized in Table~\ref{table:Performance Comparison under different partial discharge conditions}. As can be seen from Table~\ref{table:Performance Comparison under different partial discharge conditions}, RF-CNN can outperform SOH-CNN and $\Delta$SOH-CNN on all four conditions. In addition, although the MAE increases as the data range decreases for all three approaches, RF-CNN maintains a better robustness (i.e., less MAE increase) compared to the other two CNN-based estimators due to richer SOH indicators.
\newcommand{\tabincell}[2]{\begin{tabular}{@{}#1@{}}#2\end{tabular}} 
\begin{table} %\footnotesize
\begin{spacing}{1.0}
\centering 
\caption{SOH estimation MAE from different CNN-based estimators under partial discharge with different DoD ranges}
\label{table:Performance Comparison under different partial discharge conditions}
\begin{tabular}{@{}ccccc@{}} 
\toprule
Method & \tabincell{c}{DoD \\ 0.1 - 0.8} &  \tabincell{c}{DoD \\ 0.2 - 0.7}  & \tabincell{c}{DoD \\ 0.3 - 0.6}  & \tabincell{c}{DoD \\ 0.4 - 0.5} \\ \midrule
SOH-CNN  &  0.99\%   &   1.28\%  &   1.61\%  &  2.92\% \\ 
$\Delta$SOH-CNN  &    1.04\%    &  1.57\%     &   1.54\%  &  2.30\% \\ 
RF-CNN      &     0.62\%      &  0.85\%     &    1.01\%   & 1.51\%    \\ \bottomrule
\end{tabular} 
\end{spacing} 
\end{table}

%%%%%%%%%%%%%%%%%%%%%%%%%%%%%%%%%%%%%%%%%%%%%%%%%%%%%%%%%%%%%%%%%%%%%%%%%%%%%%%%%%%% 
%%%%%%%%%%%%%%%%%%%%%%%%%%%%%%%%%%%%%%%%%%%%%%%%%%%%%%%%%%%%%%%%%%%%%%%%%%%%%%%%%%%%
\section{Conclusions and Future Work} \label{section.7}
In this paper, we consider the SOH estimation problem for a single battery cell under partial discharge and propose RF-CNN as a solution. Two CNNs are used simultaneously to extract the indicators correlating to SOH and change of SOH between two consecutive discharge cycles ($\Delta$SOH) from partial discharge curves. Based on the outputs from the CNNs, a random forest model is then designed to integrate two CNNs to produce the final SOH estimate. Evaluation of the proposed approach is performed based on the partial discharge data with different DoD ranges created from a fast-discharging dataset. By comparing RF-CNN with RF-ICA, SOH-CNN, and $\Delta$SOH-CNN, enhanced accuracy and robustness are verified for the proposed approach. The sensitivity analysis of the CNN and random forest models further validates that richer indicators can be extracted by RF-CNN for SOH estimation.

In this work, the proposed approach is evaluated on a fast discharging dataset with identical 4 $C$ discharge rate under $\SI{30}{\celsius}$ ambient temperature. However, in real applications of Li-ion batteries, it is likely that SOH estimation needs to be performed with data collocated during various charge/discharge rates or ambient temperatures. Thus, our future work will focus on evaluating the effectiveness of the proposed approach on various operating conditions. In addition, since Li-ion cells are used in strings for most applications, using the proposed approach to estimate the overall battery string SOH will also be our focus in the future.

%%%%%%%%%%%%%%%%%%%%%%%%%%%%%%%%%%%%%%%%%%%%%%%%%%%%%%%%%%%%%%%%%%%%%%%%%%%%%%%%

\bibliographystyle{IEEEtran}
\bibliography{Reference}

%%%%%%%%%%%%%%%%%%%%%%%%%%%%%%%%%%%%%%%%%%%%%%%%%%%%%%%%%%%%%%%%%%%%%%%%%%%%%%%%

\end{document}